\documentclass{article}

\usepackage{arxiv}

\usepackage[utf8]{inputenc} 
\usepackage[T1]{fontenc}    
\usepackage{hyperref}       
\usepackage{url}            
\usepackage{booktabs}       
\usepackage{amsfonts}       
\usepackage{nicefrac}       
\usepackage{microtype}      
\usepackage{lipsum}
\usepackage{amsmath}
\usepackage{graphicx}
\usepackage[super]{nth}
\usepackage{xcolor}
\usepackage[title]{appendix}
\usepackage{placeins}

\newcommand{\ie}{\emph{i.e.}, }

\title{\textbf{Characterizing Twitter users behaviour during the Spanish Covid-19 first wave}}

\author{
  Bernat Esquirol \\
  Universitat de Barcelona \\
  UBICS\\
  \AND
  Luce Prignano \\
  Universitat de Barcelona \\
  UBICS\\
  \And
  Albert Díaz-Guilera \\
  Universitat de Barcelona \\
  UBICS\\
  \And
  Emanuele Cozzo \\
  Universitat de Barcelona \\
  UBICS\\
  Universitat Oberta de Catalunya\\
  CNSC/IN3 \\
  \texttt{ecozzo@uoc.edu} \\
}

\begin{document}
\maketitle
\begin{abstract} \normalsize

People use Online Social Media to make sense of crisis events. A pandemic crisis like the Covid-19 outbreak is a complex event, involving numerous aspects of the social life on multiple temporal scales. Focusing on the Spanish Twittersphere, we characterized users activity behaviour across the different phases of the Covid-19 first wave. 

Firstly, we analyzed a sample of timelines of different classes of users from the Spanish Twittersphere in terms of their propensity to produce new information or to amplify information produced by others. Secondly, by performing stepwise segmented regression analysis and Bayesian switchpoint analysis, we looked for a possible behavioral footprint of the crisis in the statistics of users' activity.

We observed that generic Spanish Twitter users and journalists experienced an abrupt increment of their tweeting activity between March 9 and March 14, in coincidence with control measures being announced by regional and State level authorities. However, they displayed a stable proportion of retweets before and after the switching point. 

On the contrary, politicians represented an exception, being the only class of users not experimenting this abrupt change and following a completely endogenous dynamics determined by institutional agenda. On the one hand, they did not increment their overall activity, displaying instead a slight decrease. On the other hand, in times of crisis, politicians tended to strengthen their propensity to amplify information rather than produce it. 

\end{abstract}


\vfill

\section{Introduction}
The COVID-19 outbreak (1 December 2019 – present) is the first global pandemic of the information age. A large-scale public debate is taking place in a hybrid media ecosystem in which online social media (OSM) in general, and Twitter in particular, play a central role \cite{GVC2020,AMA2020,shu-feng2021}. The online debate addresses all the aspects of the pandemic induced crisis that the public opinion considers to be relevant: from the origin of the virus, to the controversy surrounding vaccination campaigns; from the necessity (or lack of it) for lockdowns and other social distancing measures, to the effectiveness of the different national health systems. These are evidence that people use Twitter to make sense of the pandemic crisis, i.e. they try to understand, through interaction with others, an ongoing event that is new, uncertain, and confusing \cite{HZ2012}. However, unlike other crisis events, as for instance earthquakes or terrorist attacks, a pandemic crisis is a complex phenomena spanning different time scales. In particular, the first wave of a new infectious disease is an even more peculiar phenomenon, being completely unseen and unexpected. We can distinguish \textit{a priori} three different phases for the unfolding of the first wave in a given region: an early phase in which the virus and information about it starts to spread, a phase in which the epidemic hits a particular region and the debate about control and mitigation measures arises, and a third phase in which these measures are finally implemented. Our hypotheses, building on some recent findings \cite{AMA2020,RAA2020}, are that online users will react differently to different phases and that different type of users will react in different moments. 

In fact, it is well known that OSM react abruptly to external and internal forces, eventually resulting in deep transformations without warning \cite{Britt2015}. This inherent volatility suggests that online behaviour will change in response to crises and the associated interventions. In this sense, OSM can serve as an early warning system for any kind of emergency worldwide, from natural disasters to terrorist acts \cite{CCS2013,FSF2019}. 

Unfortunately, this useful property comes with an important drawback. The efforts made in the last few years to measure and characterize human and human-like activity in OSM mainly consist in the search for characteristic patterns in users activity data. For instance, it has been shown that some aspects of human communication are described by universal statistics which may help distinguish among different types of human and human-like users \cite{GFT2019,GKS2014}. Nonetheless, when an emergency occurs, all the typical patterns of activity are probably altered and usual classification algorithms are no longer reliable. This is even more important if we take into account the misinformation processes that accompany an ongoing crisis event, named ‘infodemics’ in the case of a pandemic \cite{GVC2020}. In order to properly identify harmful activity in the context of a crisis, either abrupt or prolonged, we need to know how different classes of users behave in such circumstances.


In this work, we characterize online collective behaviour throughout the different phases of the first wave of Covid-19, and, in particular, in response to the prevention and control measures adopted, using Spain Twittersphere as a case study. In this way, we will be able to exploit the OSM intrinsic volatility to analyze at the same time the effect of the different phases of the crisis on the collective behaviour and the long term effects of that on the typical patterns of activity for the duration of the crisis. 

To this end, we consider the whole tweeting activity, regardless of the topic of the content, instead of only tracking tweets directly related to the Covid-19 outbreak. We are thus able to compare online behaviors before and during the crisis from an overall perspective. We partition the Spain Twittersphere in three different classes of users: politicians, journalists, and generic users, since we consider that they provide a good representation of the  division of roles in a networked public sphere \cite{Auss2013}. 

\begin{figure}[h]
  \centering
    \includegraphics[width=\textwidth]{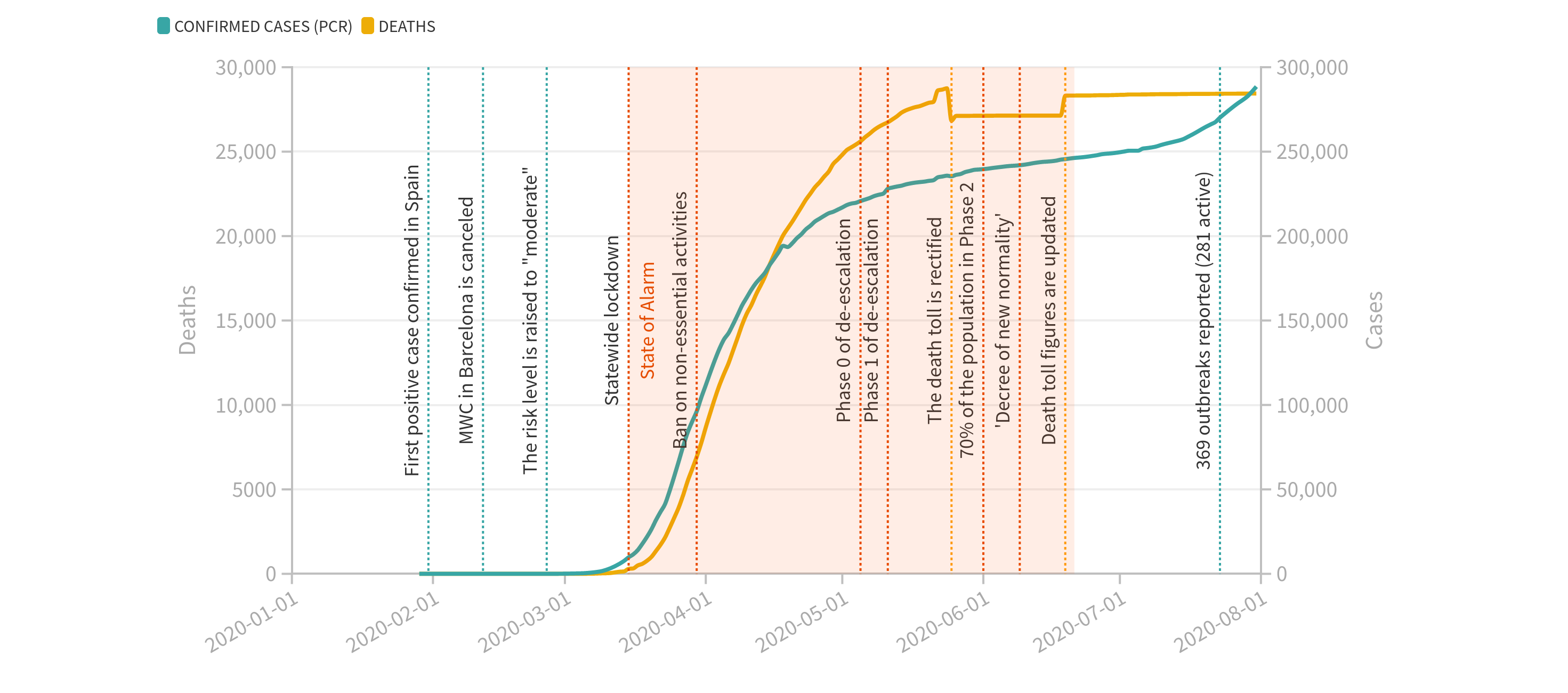}
  \caption{Timeline of Covid-19 pandemic related events in Spain: restriction measures, confirmed cases and deaths during the first wave. A interactive version of the infographic is available at \url{https://public.flourish.studio/visualisation/4432020/ }}
\end{figure}

\section{Methods}
\subsection{Data Collection}

We collected Twitter users’ timelines, \ie the time stamped list of tweets (including retweets) published by a user, of three different types of Spain-based profiles: journalists, politicians, and random users. In the following, we describe the methods used for the construction of the three set of users.

\textbf{Journalists’ accounts} 
We observed that, usually, the news media follow on Twitter the journalists working for them and journalists usually have some variation of the word journalist in their bio. Leveraging on these two elements, in order to sample Spanish journalists’ Twitter profiles, we searched the Twitter accounts of the top news media in Spain (list in appendix A) and got their friends (users being followed by the media account). Among them, we retained those who have in their bio some version of the word journalist. In this way, we obtained 1722 journalists’ profiles.

\textbf{Politicians’ accounts}
We picked the members of the \textit{Congreso XIV Legistlatura} list that contains all the members in the Spanish congress during the February to June 2020 period. There are 350 congresspersons in Spain, but not all of them have a Twitter account: altogether, we retrieved 261 politicians profiles.

\textbf{Random users}
Getting unbiased random samples of Twitter users for a given location is not an easy task given that geolocated tweets are sparse. We developed a method with the aim to reduce as much as possible the bias toward highly active users.

Our method works as follows:

\begin{enumerate}
    \item We got the 100 most recent tweets from users located in Spain through the Twitter Standard Search API 
    \item We filtered those 100 tweets so that the creator of the tweet ($u_0$) had more than a given threshold number of followers, friends, and actions (55,95, and 1000 respectively). 
    \item We got one follower and one friend of $u_0$ at random.
\end{enumerate}
We repeated the process until we retrieved around 8.000 profiles in each random set. 

Step 2 is to filter out bots or extremely inactive users. Step 3 is to avoid a bias towards high frequency users that we would have by taking user $u_0$ directly.  In this way, we get two types of random users: random-friends are those that we get sampling the friends of $u_0$, and random-followers those obtained by sampling the followers. The first group is biased toward users following many other accounts, and the other is biased towards users having many followers. We used both sets separately and compared the results of the analyses performed on them at each stage. The underlying idea is that, whenever we identifies a trait common to both random-followers and random-friends, such a trait can be regarded as a property of random users in general.

The Twitter Search API allows us to retrieve 3200 tweets for each timeline, starting from the most recent one. Hence, depending on the user's activity, we were able to cover a more or less large time period. The first data retrieval was conducted on April 28, then we periodically collected new tweets from the sampled users via the Twitter Search API. 

\subsection{Analysis of the activity}
We characterized the the activity of an individual user both from an overall (time-aggregated) perspective and analysing its evolution through time, during the different phases of Covid-19 first wave.

First of all, we characterized individual profiles based on two variables. On the one hand, we compared their tendency to publish their own contents and to share other people's tweets; on the other hand, we analysed how distributed or concentrated their retweets are, in terms of the accounts that receive them (sources). In this way, we could highlight differences between the four samples of users under study, thus tracing a preliminary behavioral profiling of the corresponding classes of accounts.

Secondly, we looked for possible abrupt jumps that the activity of a user may have experienced in particular days. To identify and quantify such jumps, we considered the mean activity before and after them, with respect to the overall mean activity, and the associate probability that a jump actually occurred that day. We performed two types of analysis and compared the results. 

\subsubsection{Production VS Amplification of information}
In this first preliminary time-aggregated analysis, we characterized each individual users according to their propensity to produce new information, \ie tweeting, or to amplify information produced by others, \ie retweeting. In addition, we quantify the diversity of the sources of the amplified information. With this aim, we introduce for each user $u$ two parameters: the fraction of replicated content $\rho_u$ and the mean entropy of the sources $h_u$, a measure of how "generalist" a given user is in terms of information/content sources. Given the number of total tweets including retweets $M_u$ emitted by a users and the number of retweets $R_u$, 

\begin{equation}
    \rho_u=\frac{R_u}{M_u}.
    \label{rho}
\end{equation}

The fraction of retweets of a given user to the same source $s$ is $p^s_u=\frac{r^s_u}{R_u}$, where $r^s_u$ is the number of retweets to the source $s$. Thus, the normalized entropy of the sources for the user $u$ is defined as

\begin{equation}
    h_u=\frac{-\sum_s p^s_u \log p^s_u}{\log R_u},
    \label{h}
\end{equation}

where the denominator is the entropy of the sources of a user that retweets each source once. 

In this way, we could compare each set of users with the others based on their distribution in the $h-\rho$ plane. In particular, we use the Kullback-Leibler (KL) divergence to measure the disparity between distributions of $\rho$ and $h$ of different set of users. The KL divergence, given by the expression

\begin{equation}
    D_{KL}(P\mid \mid Q)=\sum_{i} p(x_i)\log (\frac{p(x_i)}{q(x_i)}),
\end{equation}
 is one of the most used disparity measures between probability distributions.
 It is a measure of the information lost when $Q$ is used to approximate $P$, being $P$ and $Q$ on the same finite support. 

\subsubsection{Stepwise segmented regression analysis} The first approach that we adopted for studying possible changes in the activity of different types of Twitter users was a stepwise segmented regression (SSR) analysis \cite{Britt2015} of the timeline of each profile. Given the list of timestamps of the tweets of a user, we optimized a piecewise linear fit that divided the timeline into segments. In this way, for each timeline we obtained a number of breakpoints -- the end points of each segment -- marking the moments of abrupt change in the tweeting frequency of the user. The optimal breakpoints were estimated by minimizing the residual sum of squares \cite{BP2003}, while their number was fixed by minimizing the Bayesian Information Criterion of the model. 

In this way, we could associate to each user $u$ the set of breakpoints $\tau^u_1, \tau^u_2,...,\tau^u_n$, \ie the days in which their activity changed abruptly, if any. Finally, we characterized users’ breakpoints by the relative jump

\begin{equation}
    J^u_i=\frac{L^u_i-L^u_{i-1}}{m^u}=\frac{\Delta L^u_i}{m^u}
\end{equation}

where $L_i^u$ is the tweeting frequency of user $u$ in the interval $[\tau^u_i,\tau^u_{i+1})$ and $m^u$ is the mean frequency on the whole user u retrieved timeline. 

We could hence describe the overall activity by summing separately positive and negative jumps:
\begin{equation}
    J^{\pm}(\tau) = \sum_u J_i^u \theta(\pm J_i^u) \delta_{\tau}^{\tau_i^u} 
\end{equation}

\subsubsection{Bayesian switchpoint analysis}

The second approach was based on Bayesian inference. We performed a Bayesian switchpoint (BS) analysis, which looks for a single moment of change in the rate at which events occurred. The model assumes, for each user, Poisson-distributed tweeting events with constant but potentially different rates, before and after a switchpoint.

\begin{center}
\begin{align}
    \lambda_1^{(0)} \ :\ \ \ & \text{Exponential (rate =$\alpha$)}\\
    \lambda_2^{(0)} \ :\ \ \ & \text{Exponential (rate =$\alpha$)}\\
    \tau \ :\ \ \ & \text{Uniform }[0,1)
\end{align}
For $i=1 \dots n$:
\begin{equation}
\lambda_i=
\begin{cases} 
    \lambda_1^{(0)} \ \ \ \text{if} & \tau \le i/n  \\
    \lambda_2^{(0)} \ \ \ \text{if} & \tau > i/n
\end{cases}
\end{equation}
\end{center}

Our prior for $\lambda_{(1,2)}$ was an exponential distribution with rate $\alpha$ while our prior for $\tau$ was a uniform distribution over the whole time span of the timeline.

An Hamiltonian Monte Carlo method \cite{neal2011mcmc} as implemented in TensorFlow \cite{abadi2016tensorflow} was used for sampling from the relevant posterior distributions of $\lambda_{1,2}$  and $\tau$ .

Thus, we could associate to each user a probability distribution for the switchpoint and a probability distribution for the two different tweeting rates, before and after the switch respectively. To carry out an overall analysis of the behavior of the Spanish Twittersphere, as for the SSR analysis, we needed to characterize the switchpoints. By approximating the distributions of the $\lambda$s with a delta centered in their mode, we obtained the following expression for the relative jump:

\begin{equation}
    J^u(\tau_i)=P(\tau^u=\tau_i)\frac{\tilde{\lambda^u_2}-\tilde{\lambda^u_1}}{m^u}=\frac{\Delta\lambda^u}{m^u}
\end{equation}

where $\tilde{\lambda}_{1,2}$ is the mode of the respective posterior distribution. 
We could then characterize the overall activity by summing separately positive and negative jumps of each user weighted by the jump probability:

\begin{equation}
    J^{\pm}(\tau)=\sum_{u}J^u(\tau_i)\Theta(\pm J^{u}(\tau_i))\delta^{\tau_i}_\tau
\end{equation}

\section{Results}

\subsection{Overall activity of different classes of users}
\begin{figure}[ht] 
  \begin{minipage}[b]{0.5\linewidth}
    \centering
    \includegraphics[width=.5\linewidth]{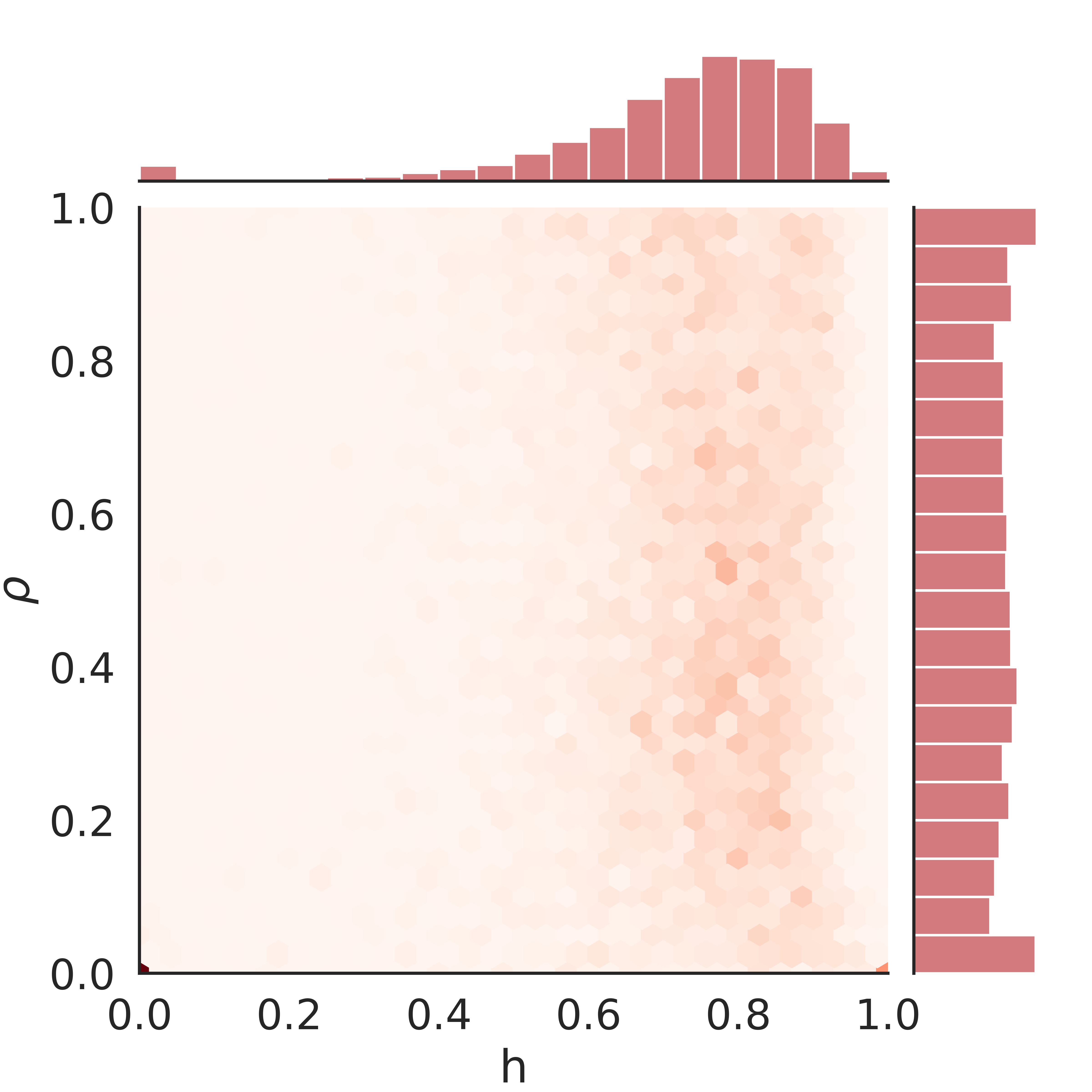} \\
    (a) Random Followers
    \vspace{4ex}
  \end{minipage}
  \begin{minipage}[b]{0.5\linewidth}
    \centering
    \includegraphics[width=.5\linewidth]{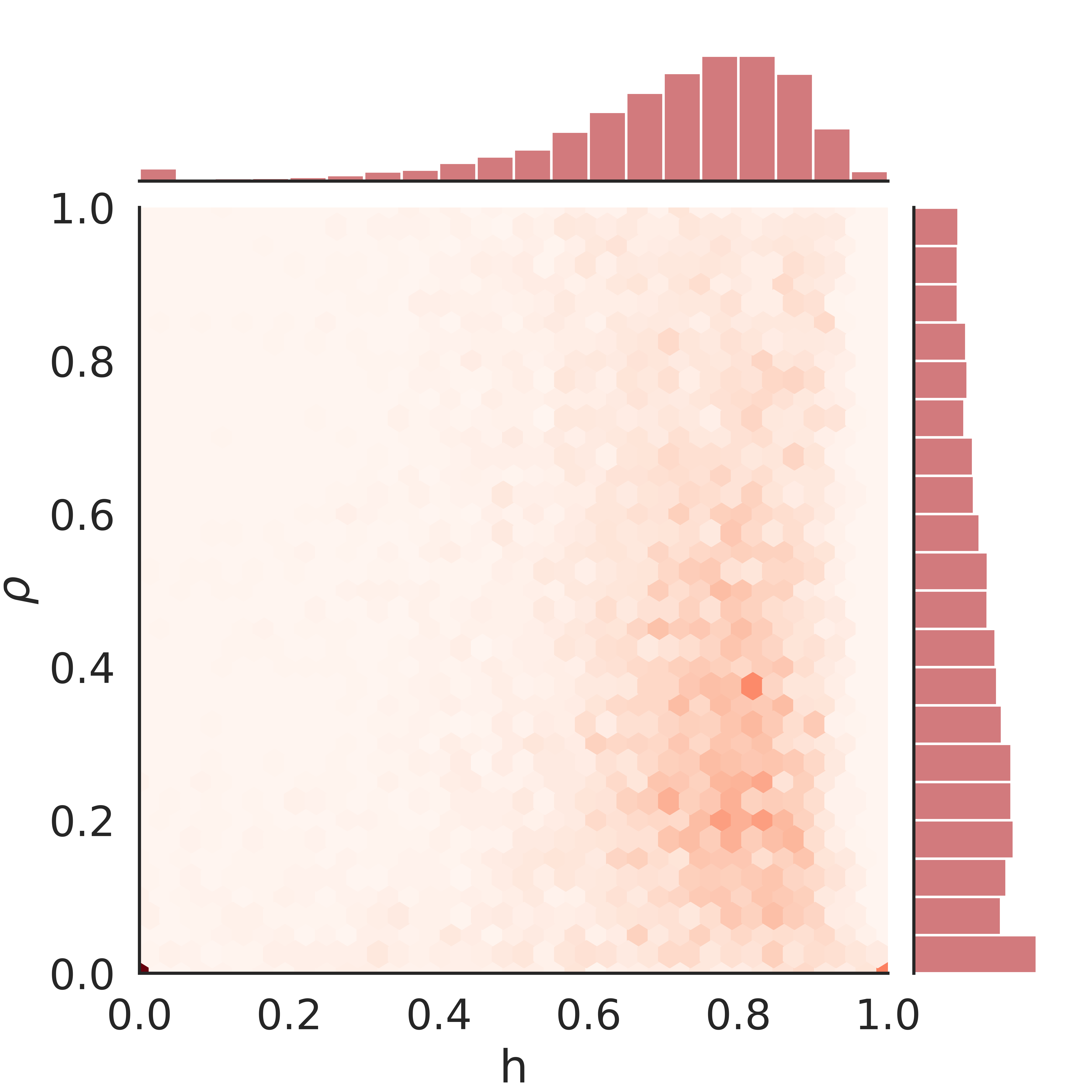} \\
    (b) Random Friends 
    \vspace{4ex}
  \end{minipage} 
  \begin{minipage}[b]{0.5\linewidth}
    \centering
    \includegraphics[width=.5\linewidth]{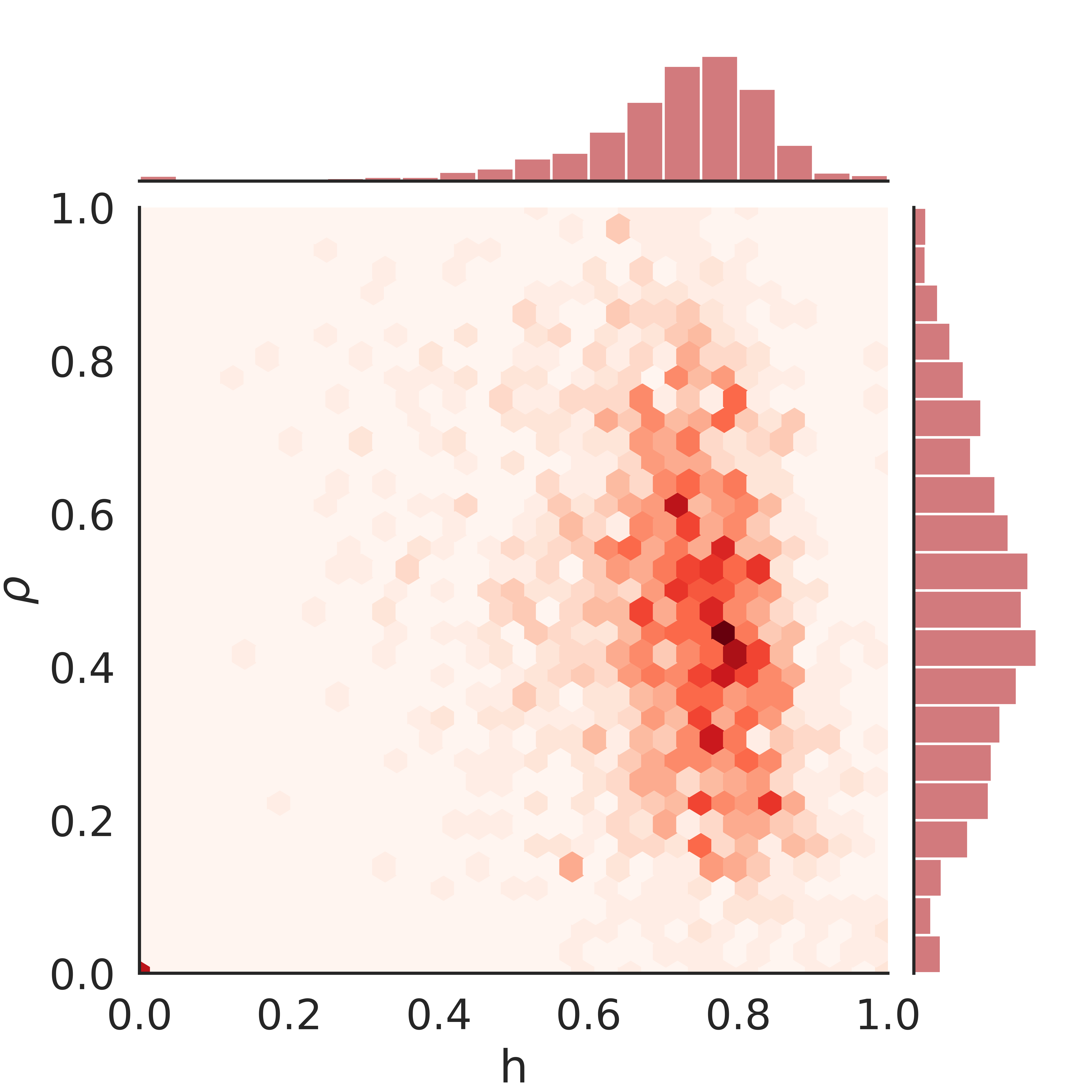} \\
    (c) Journalists
    \vspace{4ex}
  \end{minipage}
  \begin{minipage}[b]{0.5\linewidth}
    \centering
    \includegraphics[width=.5\linewidth]{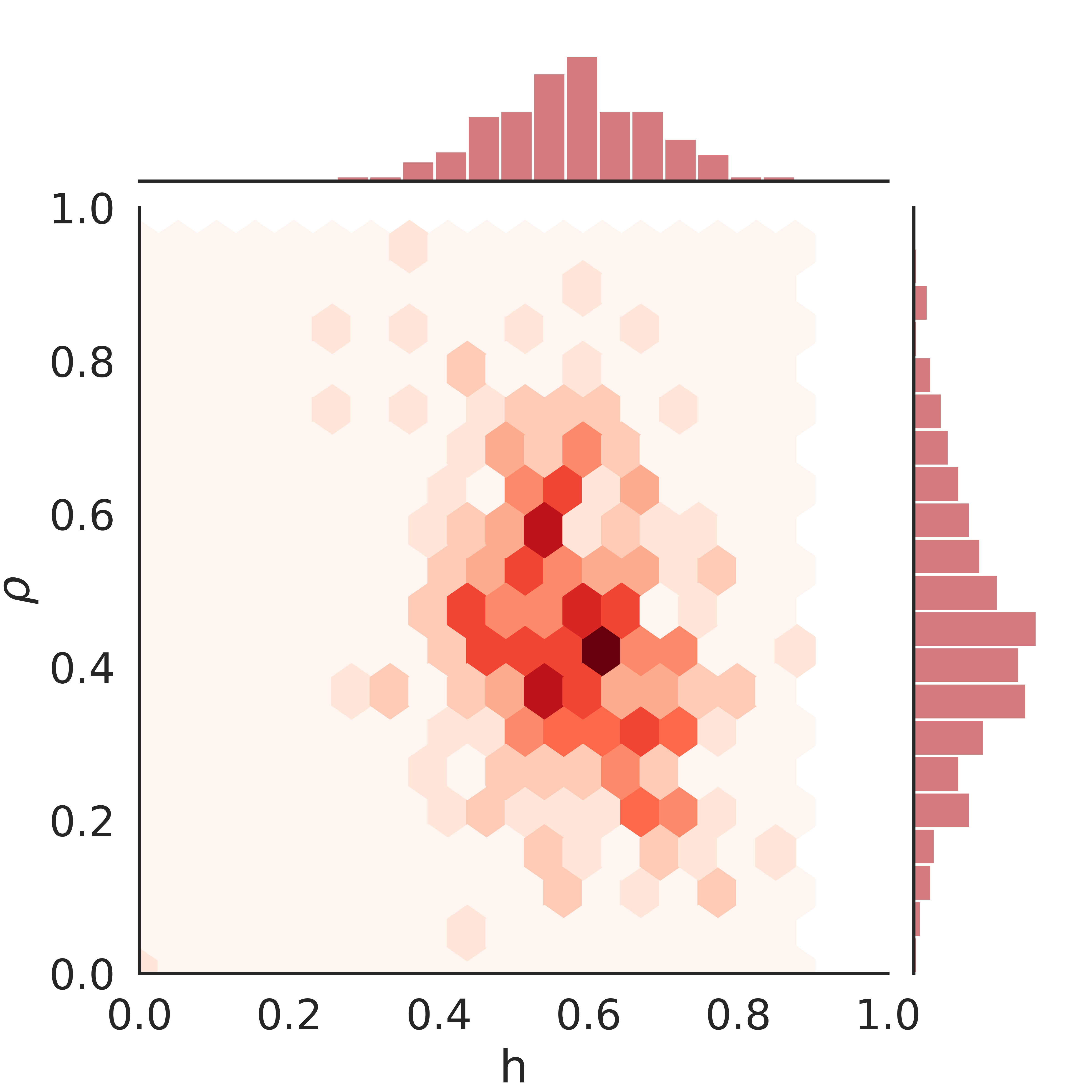} \\
    (d) Politicians 
    \vspace{4ex}
  \end{minipage} 
 \caption{Distribution of the fraction of retweets $\rho$ and normalized entropy $h$ for each one of the considered classes of users.}
  \label{figrho-h} 
\end{figure}

Comparing the different sets of users with respect to their tendency to produce new information rather than amplify already existing contents ($\rho$) and how they distributed their retweets among different sources ($h$), we observed that the two sets of random profiles -- followers and friends -- were practically indistinguishable (Fig.\,\ref{figrho-h}, upper panels). They both showed a nearly flat distribution of $\rho$ (vertical axe), covering equitably all the spectrum from pure replicator ($\rho=1$) to pure content producer ($\rho=0$). For what concerns $h$ (horizontal axe), both groups displayed a markedly right skewed distribution, a clear indication that most of them tended to distribute their retweets almost uniformly among different sources. Journalists (Fig.\,\ref{figrho-h}, lower right panel) appeared to be quite similar to random users, but avoiding extreme values of $\rho$, in line with the expectation that a journalist will always tend to retweet her media and colleagues as well as produce new contents. Finally, politicians (Fig.\,\ref{figrho-h}, lower left panel) displayed the most peculiar activity patterns. They are more similar to journalists in $\rho$ but showing a complete different distribution of $h$ (symmetric, with the mean and the mode falling slightly above $0.5$).

\begin{figure}[ht] 
  \begin{minipage}[b]{0.5\linewidth}
    \centering
    \includegraphics[width=.7\linewidth]{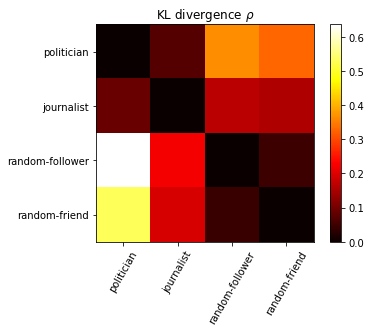} \\
   KL divergence for $\rho$ distributions
    \vspace{4ex}
  \end{minipage}
  \begin{minipage}[b]{0.5\linewidth}
    \centering
    \includegraphics[width=.7\linewidth]{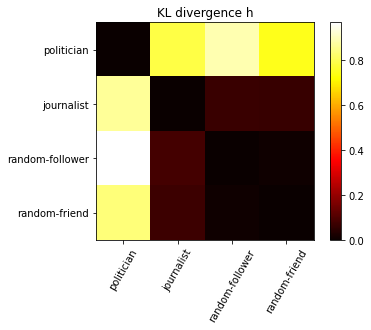} \\
    KL divergence for $h$ distributions
    \vspace{4ex}
  \end{minipage} 
  \caption{Kullback-Leibler (KL) divergence of $\rho$ (left panel) and $h$ (right panel) distribution between classes of users.}
  \label{figKL} 
\end{figure}

Such differences and similarities are quantitatively described by the KL divergence for the distributions of $\rho$ and $h$, represented as heatmap matrices in Fig.\,\ref{figKL}.

\FloatBarrier
\subsection{Time evolution}
Once we clarified that random users differed from politicians and, although to a lesser extent, journalists in terms of their overall (re)tweeting activity, the next step was to compared the evolution of their activity rate during the first wave of Covid-19.

\begin{figure}[htp]
  \centering
    \includegraphics[width=0.7\textwidth]{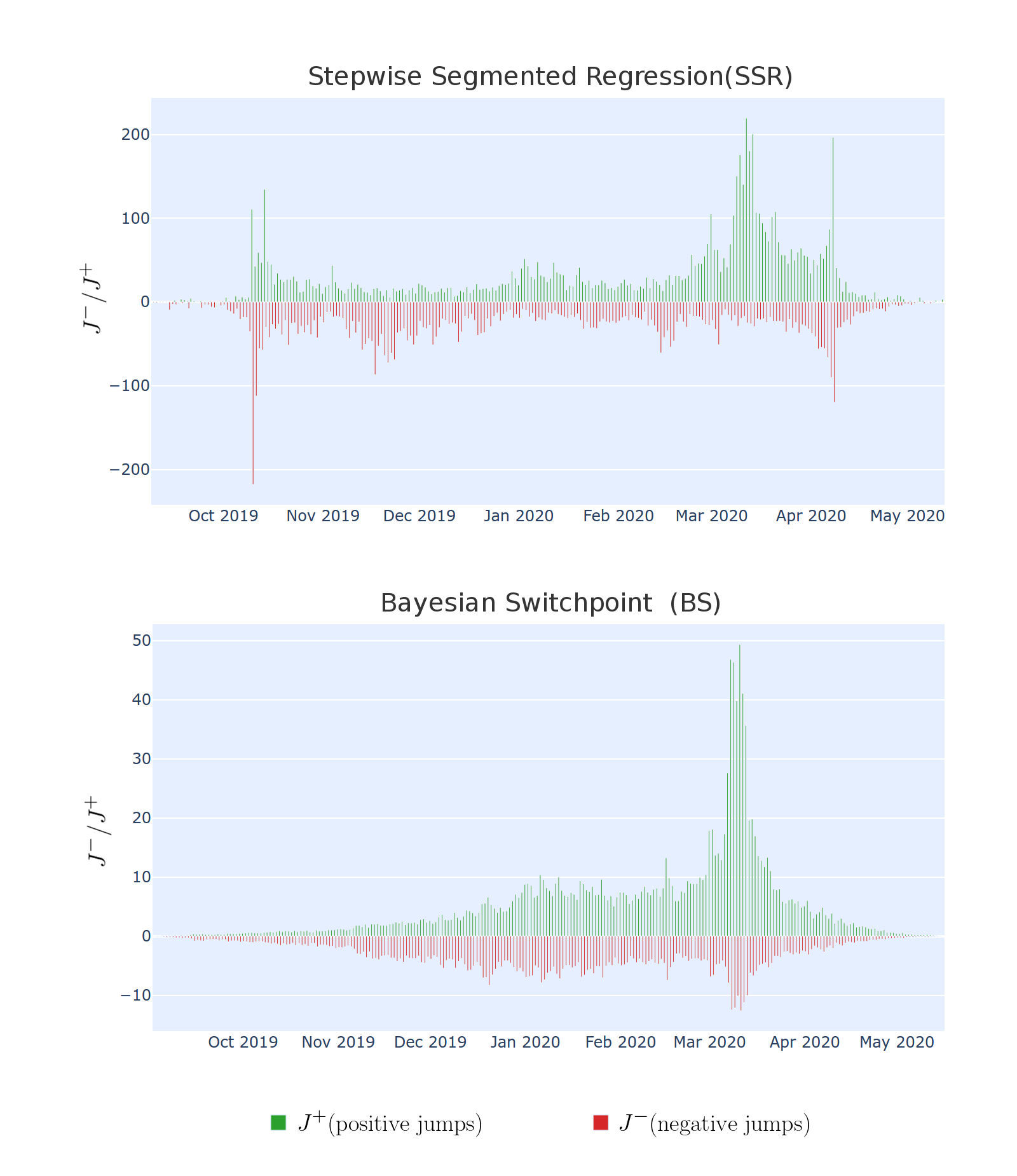}
  \caption{Ratio of the total daily negative activity jump to the total daily positive activity jump calculated by performing stepwise segmented regression (SSR, upper panel) and Bayesian switchpoint analysis (BS, lower panel) for the whole set of users.}
\label{SSR-BS}
\end{figure}

For a preliminary general inspection, we considered all the users together, regardless their class. From both the SSR and BS analyses (Fig.\,\ref{SSR-BS}), we obtained a clear indication that in the week between March 9 and March 14 the majority of Spanish twitter users abruptly changed their tweeting behaviour, starting to tweet with higher frequency. These days coincided with the first control measures being introduced at regional level on Monday 9 and the declaration of the state of alarm by the prime minister of Spain on Friday 13, which implied enforcement of a state level lockdown. 

The BS analysis also detected a noticeable decrease in the tweeting activity in those days, but much less intense than the opposite behaviour.

As a final note, the SSR and BS analyses were discordant near the boundary of the considered time window. However, this fact can be easily explained considering that these two models are biased in opposite ways in these regions. The analysis had hence to be regarded as inconclusive for what concerns both the end of 2019 and May 2020.

\begin{figure}[htp]
  \centering
    \includegraphics[width=0.9\textwidth]{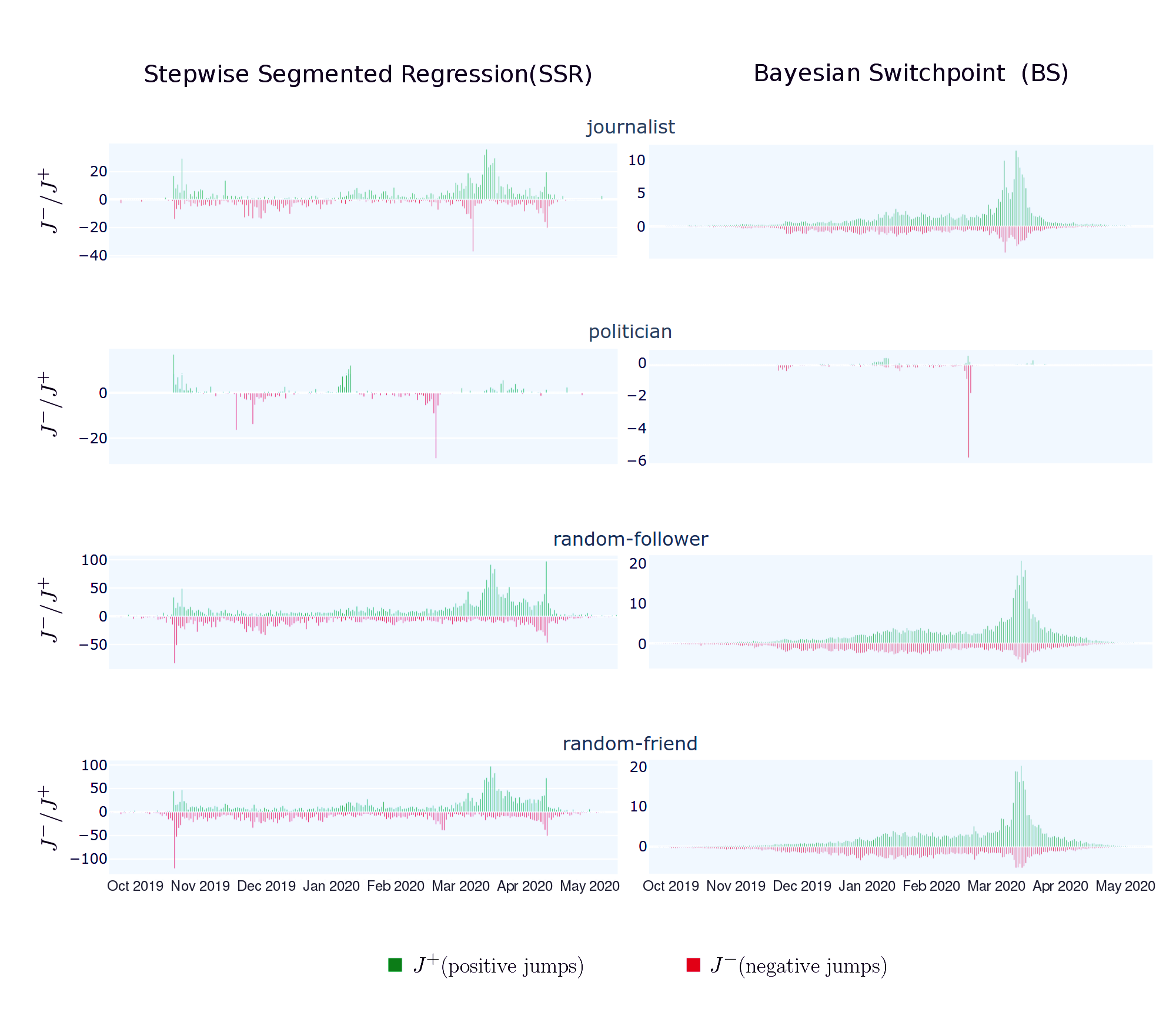}
  \caption{Ratio of the total daily negative activity jump to the total daily positive activity jump calculated by performing stepwise segmented regression (SSR, upper panels) and Bayesian switchpoint analysis (BS, lower panels) for each individual class of users.}
\label{SSR-BS-sets}
\end{figure}

We repeated the analyses for the four sets of profiles separately (Fig.\,\ref{SSR-BS-sets}). In this case,  we could observe that the politicians were the only group that did not display any change in their activity in the week of the lockdown declaration. Contrarily, both analyses indicate a positive jump in the activity near January 7, the day on which the two parties forming the coalition government found an agreement, as well as a pronounced negative jump near February 12, day in which the composition of the government was announced by the prime minister. Finally, journalists also showed an interesting behaviour. Relative positive jumps concentrate in the aforementioned week, but they also presented a secondary peak on March 2, the day before the first Covid-19 death was officially confirmed in Spain.

\FloatBarrier

\begin{figure}[ht] 
  \label{figrho-hba} 
  \begin{minipage}[b]{0.5\linewidth}
    \centering
    \includegraphics[width=.8\linewidth]{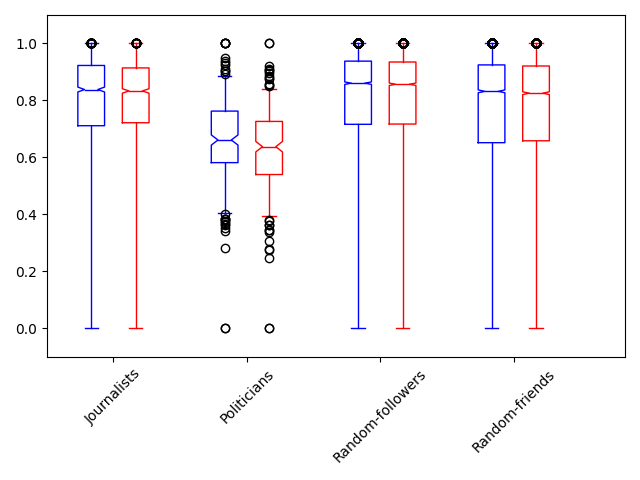} 
    \vspace{4ex}
  \end{minipage}
  \begin{minipage}[b]{0.5\linewidth}
    \centering
    \includegraphics[width=.8\linewidth]{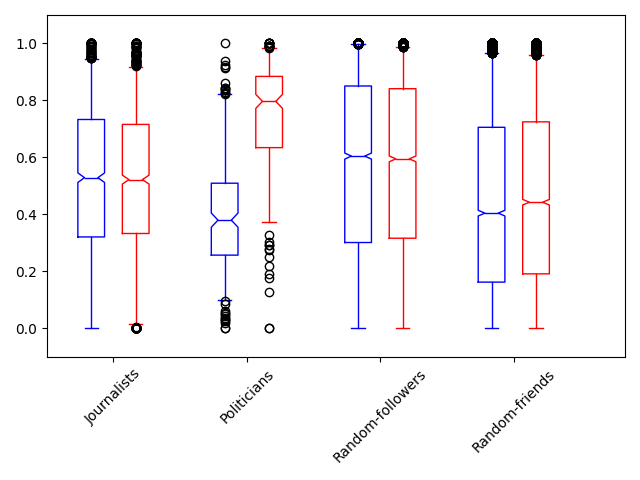} 
    \vspace{4ex}
   \end{minipage}
   \caption{
    $h$ (left panel) and $\rho$ (right panel) statistics before (blue) and after (red) March 9 of, from left to right, journalists, politicians, random followers, and random friends. Whiskers are set at the 5th and 95th percentiles. Dots represent outliers. Notches represent 95\% CI of the medians. Median values obtained by bootstraping data 1000 times.}
    \label{h-rho-b-a}
\end{figure}

Finally, we compared the distribution of $h$ and $\rho$  for the four classes of users before and after March 9 and starting on the first of January (Fig.\,\ref{h-rho-b-a}). Contrary to what we observed for the activity, the only type of users that change their behaviour are politicians. In particular, we observed a steep increase in their propensity to amplify information produced by others instead of publishing their own content. 

\FloatBarrier
\section{Conclusions}

In this work, we have shown how a statistical analysis of individual users' activity on online social media can highlight the relation between crisis events and abrupt changes in behaviours. In particular, our results indicate that the debate about the government of the crisis triggers collective participation more than events itself. This is in line with findings in \cite{RAA2020}, in which the authors demonstrated a robust large increase in the use of words related to control measures only since the first week of March. Generic Covid-related words, instead, started to be increasingly used since January.

The analysis of the individual activity of users in the Spanish Twittersphere shows that an abrupt change occurs for a large part of them in coincidence with the days on which the control and mitigation measures were announced in Spain, at regional and State level. In this sense, we can claim that it is not the epidemic outbreak that drives the change in the users’ activity. Indeed, the main driver is rather the political reactions to the crisis. Interestingly enough, we do not observe any major activity change at the beginning of the lockdown period, suggesting that there is no further effect on the online activity of the change in the physical social behaviour. The only observable consequence of the enforcement of the “Stay at home” recommendations is precisely the absence of a second change, which under more ordinary circumstances would have reestablished the overall activity at the previous level. Differently from the case of a usual highly participated political debate, the discussion around the introduction of the state of alarm and the related measures, determined a persistent increase in the online activity. 

It is worth mentioning that the phase in which we observe major changes in users activities roughly matches the transition between two different phases of the associated social epidemics \cite{Strong90} as shown in \cite{Aiello2020}. In particular, extrapolating results from \cite{Aiello2020}, jumps in activities should be located between the suspended reality phase, in which people express anger about the looming feeling that things were about to change, and the acceptance phase, started after the authorities imposed physical-distancing measures, in which people adjust to a "new normality". This observation calls for a deeper understanding of the relation between online activity and psychological states during a crisis. 

Additionally, we have been able to collect evidence that points to the fact that politicians' activity has a dynamics that is mostly endogenous, while journalists anticipate trends. Politicians did not follow the general pattern described above. On the contrary, their activity responds rather to the institutional agenda itself. With respect to the type of activity -- production or amplification of contents --, only politicians change their attitude, while for other classes, the increased activity is not accompanied by a change in the type of activity. Summarizing, policymakers tended to suppress their individual voices, likely in the attempt to provide a more coherent view on the topic, while reducing their involvement in other debates. On the contrary, ordinary citizens probably just added a new persistent theme to their usual activity, without changing quantitatively their content consumption habits. Among them, journalists took an expected leadership role.

\newpage

\begin{appendices}
\section{List of news media twitter handlers}
el\_pais, JotDownSpain, eldiarioes, elespanolcom, revistamongolia, la\_ser, \_infoLibre, EFEnoticias, elmundoes, elconfidencial, indpcom, ctxt\_es, publico\_es, ondacero\_es, cuatro, LaVanguardia, europapress, laSextaTV, rtve
\section{Overall activity}
\begin{figure}[h]
  \centering
    \includegraphics[width=0.5\textwidth]{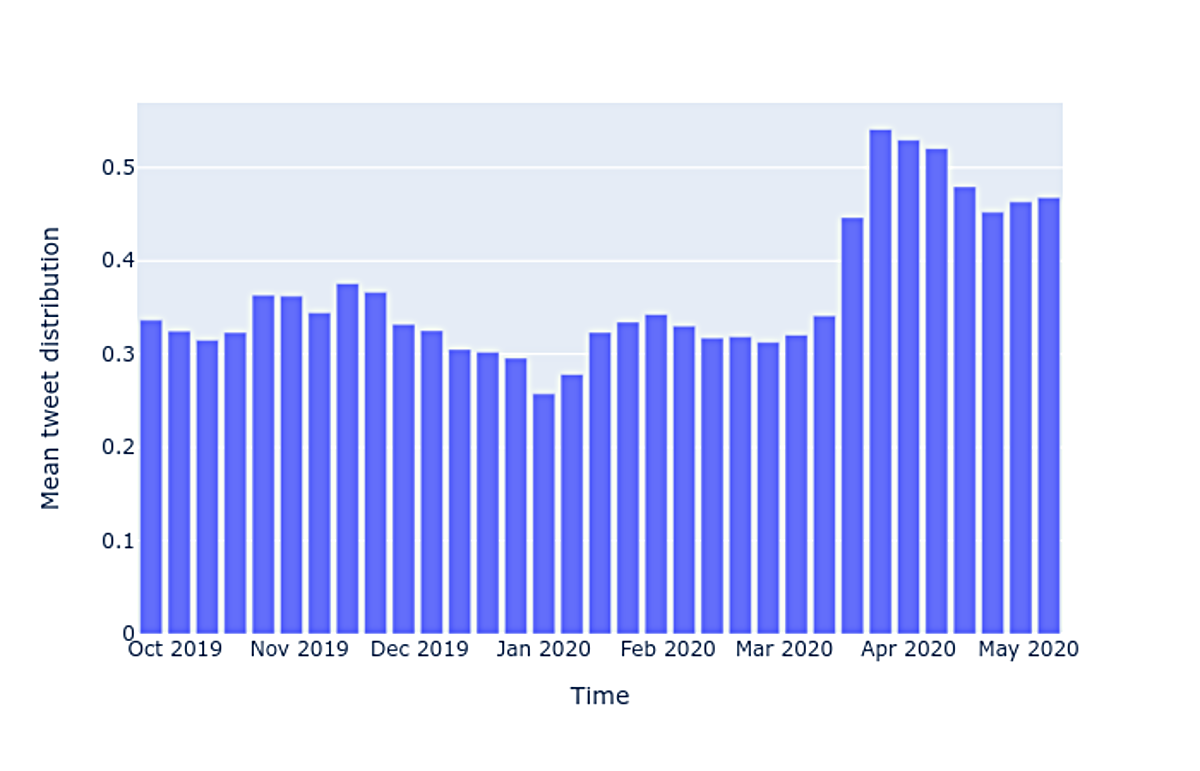}
  \caption{Activity per week for all our users.}
\label{afig1}
\end{figure}

In fig. \ref{afig1}, we represent the mean number of tweets per day for each week. We observe that the overall activity increases in the week after the one in which we observe the majority of individual changes. 

\section{Distributions of the number of breakpoints in the Stepwise Segmented Regression Analysis}

\begin{figure}[h]
  \centering
    \includegraphics[width=0.7\textwidth]{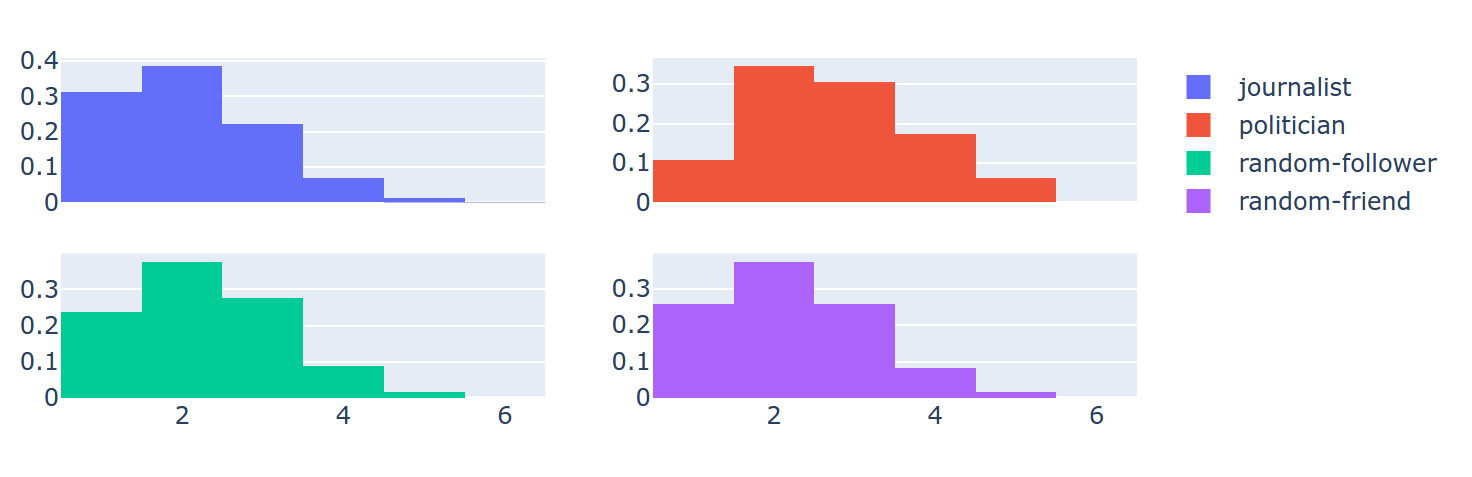}
  \caption{Number of breakpoints for each user type.}
  \label{afig2}
\end{figure}

In fig. \ref{afig2}, we show the fraction of the number of optimal breakpoints for each user type. We can see that the distribution is quite similar among different user types. We can observe that the maximum of the distribution is always centered at 2 breakpoints, and that few users reach the 5 breakpoints. This serves as an \textit{a posteriori} validation of the maximum number of breakpoints we set in our model.

\section{Bayesian Switchpoint Analysis in presence of multiple switches.}

Users timeline may show more than one switch, while our model assumes there is just one. We show some examples of the results we get when synthetic timelines are generated with multiple switches. 

\begin{figure}[h!]
  \centering
    \includegraphics[width=0.7\textwidth]{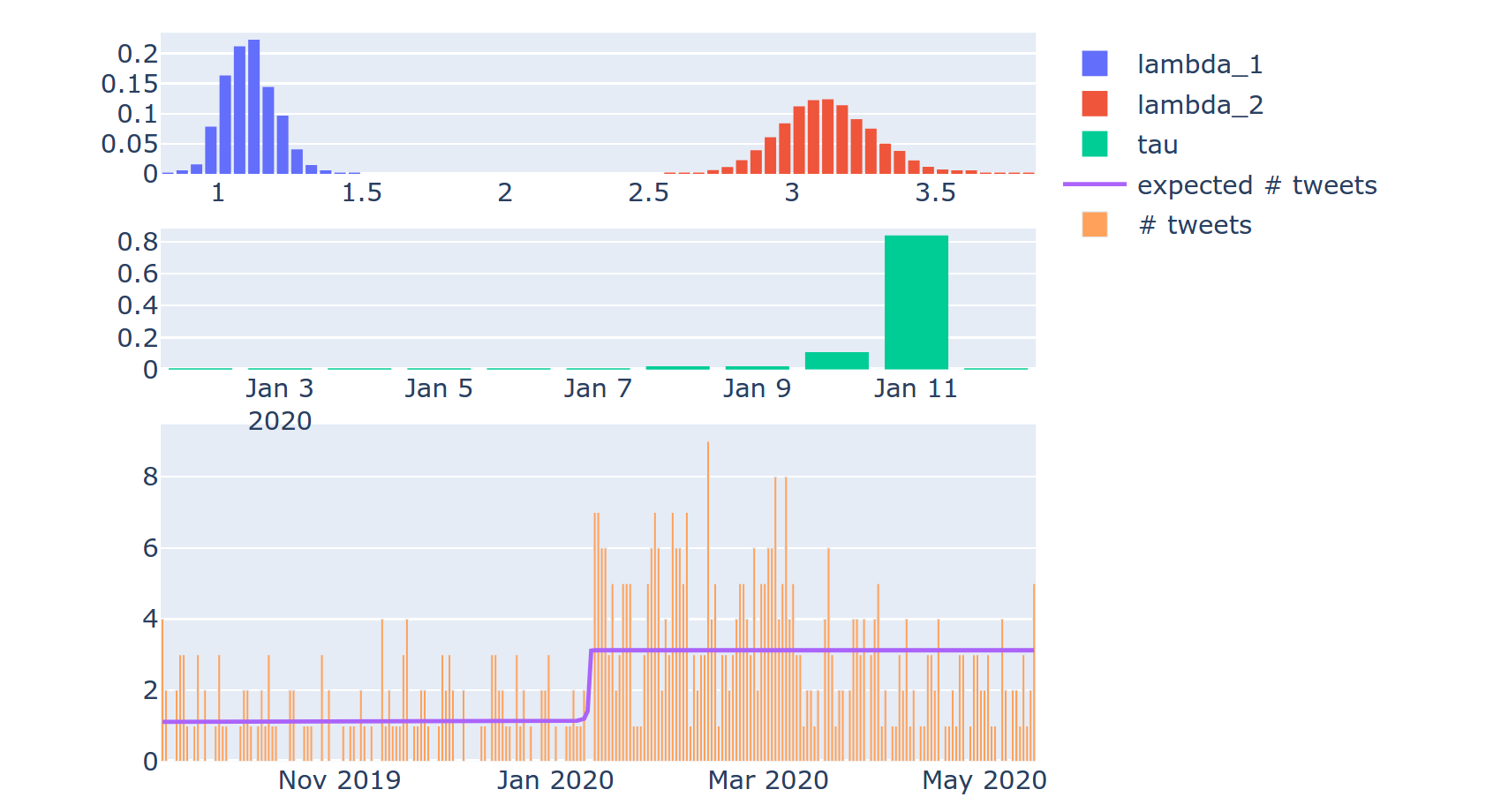}
  \caption{Syntetic data with $\lambda$ changing from 1 to 5 at the 2019-01-12 and from 5 to 2 at the 2020-03-02. For the model we obtain a p-value=0.46}
\end{figure}

\begin{figure}[htp]
  \centering
    \includegraphics[width=0.7\textwidth]{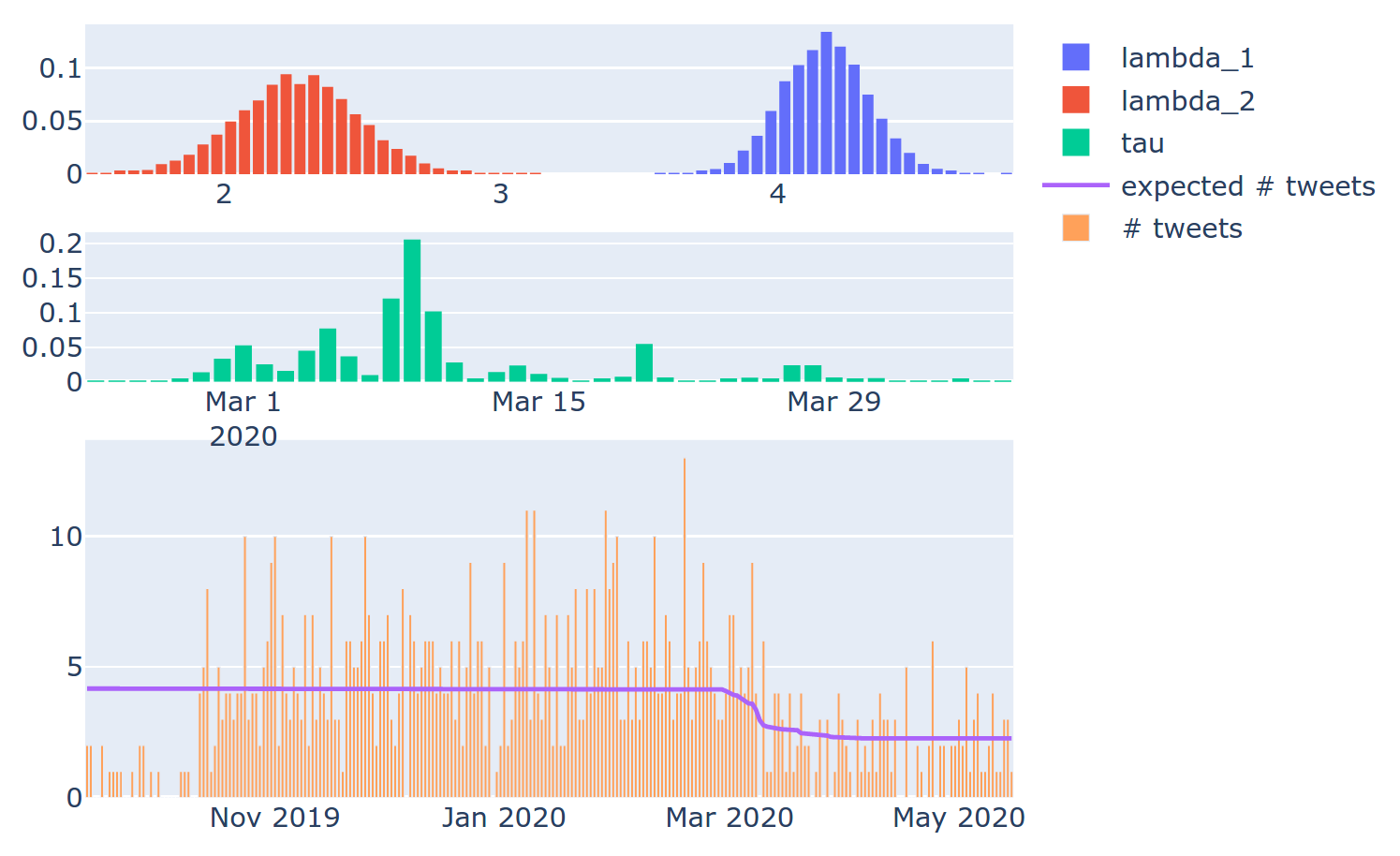}
  \caption{Syntetic data with $\lambda$ changing from 1 to 5 at the 2019-10-12 and from 5 to 2 at the 2020-03-02. For the model we obtain a p-value=0.50}
\end{figure}

In both examples we can see how in case of multiple switches the model will obviously get only one of them. We can conclude that we can use the p-value as a way of knowing if there are some existing breakpoints in the data that we missed. When the p-value is low, it does not mean that the model has to be rejected, but rather that it is choosing one of the breakpoints present in the data, while other breakpoints probably exist.

\section{Bayesian Switchpoint Analysis with non-Poissonian models.}

For comparison, we also analysed other models for a given example. 

\subsection{Sigmoidal}

We analyze here the sigmoidal model. Instead of a sudden switch, we now have a smooth transition between the two $\lambda$s. While in this way we can reduce the uncertainty in $\tau$, in general the two models give very similar results. 

\begin{figure}[htp]
  \centering
    \includegraphics[width=0.7\textwidth]{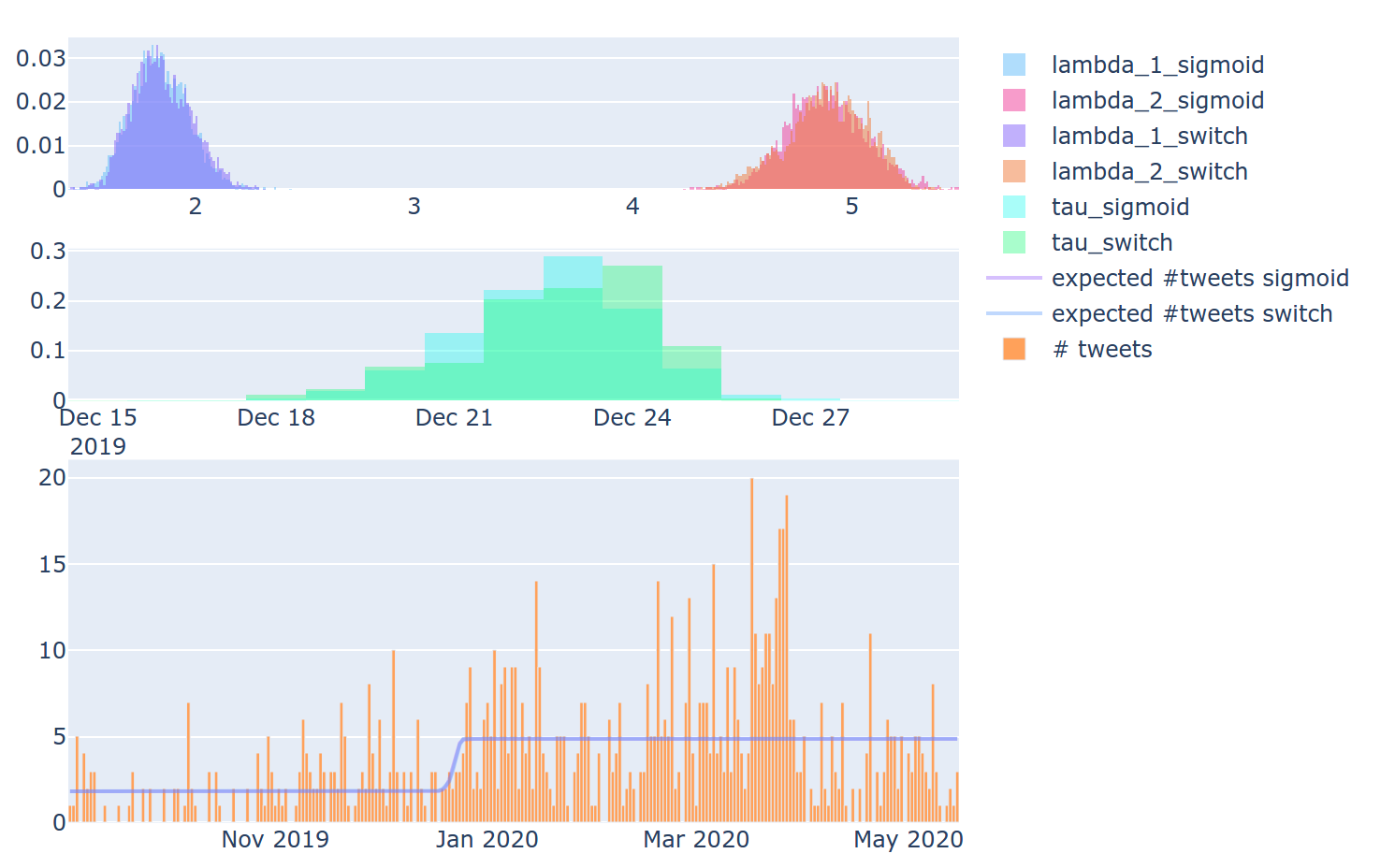}
  \caption{Comparison between Poissonian and Sigmoidal model}
\end{figure}

\subsection{Multistate model}

As in the case of the breakpoints model in SSR, we can allow for multiple states or levels represented by more than two $\lambda$s. The maximum number of levels being ($L_{\text{MAX}}$).

We will have several realization of the model with different ($L_{\text{MAX}}$) and get the optimal model minimizing the Bayesian Information Criterion.

\begin{figure}[ht]
  \centering
    \includegraphics[width=0.9\textwidth]{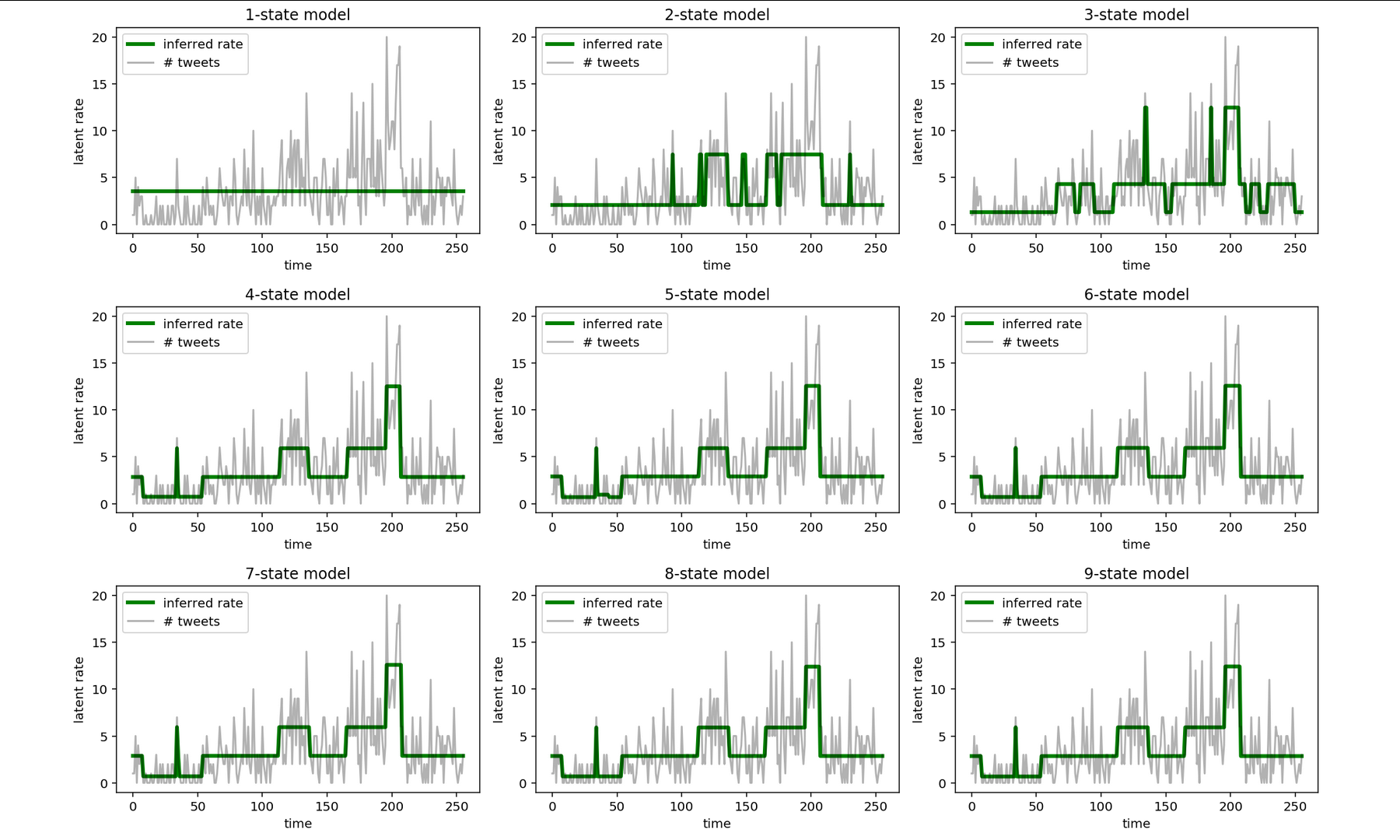}
  \caption{Multistate model for different states}
  \label{afig7}
\end{figure}

This model has the advantage of fitting the data more accurately, but the computation time and the simplicity of the result are worse than that of the linear breakpoints model. For example, we can see in fig. \ref{afig7} that the 4-state model has a small spike at a given day, this is useful to detect anomalies in the activity, but when we look for changes that persist through time, the linear breakpoints model introduces less noise.
\end{appendices}
\FloatBarrier

\bibliographystyle{unsrt}
\bibliography{references}

\end{document}